
\magnification=\magstep1
\hsize=15.7truecm \vsize=23.4truecm
\baselineskip=6mm
\font\bg=cmbx10 scaled 1200
\footline={\hfill\ -- \folio\ -- \hfill}
\def\prenum#1{\rightline{#1}}
\def\date#1{\rightline{#1}}
\def\title#1{\centerline{\bg#1} \vskip 5mm}
\def\author#1{\centerline{#1} \vskip 3mm}
\def\address#1{\centerline{\sl#1}}
\def\abstract#1{{\centerline{\bg Abstract}} \vskip 3mm \par #1}

\def\references{{\centerline{\bg References}} \vskip 3mm}

\def\chapter#1{\centerline{\bg#1} \vskip 5mm}
\def\section#1{\centerline{\bg#1} \vskip 5mm}
\def\endpage{\vfill \eject}

\fontdimen5\textfont2=1.2pt

\prenum{KOBE--92--02}
\date{February 1992}

\vskip 10mm

\title{The General Class of String Theories on Orbifolds}

\vskip 25mm

\author{Makoto SAKAMOTO and Masayoshi TABUSE}

\address{Department of Physics, Kobe University}
\address{Nada, Kobe 657, Japan}

\vfill

\abstract{ We investigate the following three consistency conditions for
constructing string theories on orbifolds: i) the invariance of the
energy-momentum tensors under twist operators, ii) the duality of
amplitudes and iii) modular invariance of partition functions. It is shown
that this investigation makes it possible to obtain the general class of
consistent orbifold models, which includes a new class of orbifold
models.}

\endpage

\section{1. Introduction}

In the construction of realistic four-dimensional string models, various
approaches have been proposed [1-8]. Among them, the orbifold
compactification [1] is probably the most efficient method and is believed
to provide a phenomenologically realistic string model. The search for
realistic orbifold models has been continued by many authors [9-12].
However, only a very small class of orbifold models has been investigated
so far and any satisfactory orbifold models have not yet been found. A
more general and systematic investigation should be required.

An orbifold [1] will be obtained by dividing a torus by the action of a
discrete symmetry group $G$ of the torus. A large number of studies have
been made on a class of orbifold models in which any group element $g$ of
$G$ is represented by [1]
$$g=(U,v),\eqno(1-1)$$
or more generally for asymmetric orbifolds [13]
$$g=(U_L,v_L;U_R,v_R),\eqno(1-2)$$
where $(U_L,U_R)$ are rotation matrices and $(v_L,v_R)$ are shift vectors.
The action of $g$ on a left- and right-moving string coordinate $
(X_L,X_R)$ is given by
$$g(X_L,X_R)g^\dagger=(U_L(X_L+2\pi v_L),U_R(X_R-2\pi v_R)).\eqno(1-3)$$
However, to the best of our knowledge, there have been few discussions
about the questions whether there might be any other class of consistent
orbifold models and whether the action of $g$ on the string coordinate
might in general be given by eq.(1-3).

The purpose of this paper is to answer the question what is the most
general class of consistent bosonic string theories on orbifolds. We shall
show that any group element $g$ of $G$ can indeed be specified by eq.(1-2)
but that the action of $g$ on the string coordinate given in eq.(1-3) is
not, in general, correct.

In section 2, we describe the basic setup and discuss consistency
conditions of string theories on orbifolds. In section 3, we investigate
the cocycle property of vertex operators and present an explicit operator
representation of cocycle operators, which are attached to vertex
operators to ensure the duality of amplitudes. In section 4, we discuss
the duality of amplitudes in detail. It is shown that the requirement of
the duality of amplitudes severely restricts the allowed action of $g$ on
the string coordinate and that the transformation (1-3) has to be modified
in general.
In this analysis, we see that the representation of the cocycle operator
given in section 3 plays an crucial role. In section 5, we discuss one
loop modular invariance of partition functions and see that this argument
justifies our prescription. In section 6, we present an example of
orbifold models, which will give a good illustration of our formalism.
Section 7 is devoted to discussions.
In appendix A, we prove a theorem, which will be used in the text. In
appendix B, we prove that any representation of cocycle operators can
reduce to the representation given in section 3 by a suitable unitary
transformation (up to a constant phase).

\vskip 10mm

\section{2. Operator Formalism for Bosonic String Theories on Orbifolds}

An orbifold [1] will be obtained by dividing a torus by the action of a
suitable discrete group $G$. In the construction of an orbifold model, we
start with a $D$-dimensional toroidally compactified closed bosonic string
theory which is specified by a $(D+D)$-dimensional lorentzian even
self-dual lattice $\Gamma^{D,D}$ [14]\footnote{$^\dagger$}{The
generalization to a lorentzian even self-dual lattice with signature $
(p,q)$ will be straightforward and will not be discussed here.}, on which
the left- and right-moving momentum $(p^I_L,p^I_R)$ $(I=1,\cdots,D)$ lies.
Since an orbifold model is given by specifying the action of each group
element $g$ of $G$ on the left- and right-moving string coordinate $
(X^I_L,X^I_R)$ $(I=1,\cdots,D)$, our aim of this paper is to answer the
question what is the most general allowed action of $g$ on the string
coordinate. To determine the allowed action of $g$ on the string
coordinate, we require the following three conditions:

\item{(i)} The invariance of the energy-momentum tensors under the action
of $g$; This condition guarantees the single-valuedness of the
energy-momentum tensors on the orbifold.
\item{(ii)} The duality of amplitudes; This is one of the important
properties of string theories [15,16].
\item{(iii)} Modular invariance of partition functions; Modular invariance
plays an important role in the construction of consistent string models
[16] and conformally invariant field theories [17]. Modular invariance may
ensure the ultraviolet finiteness and the anomaly free condition of
superstring theories [16,18]. The space-time unitary also requires modular
invariance [19].

Although the first and the third conditions (i) and (iii) have already
been considered, little attention has been given to the second condition
(ii) so far. As we will see later, our main results will be obtained from
the detailed analysis of the second condition (ii).

Let us first consider the condition (i), that is, the energy-momentum
tensors have to be invariant under the action of $g$. The energy-momentum
tensors of the left- and right-movers are given by
$$T_L(z)=\lim_{w\to z}{1\over 2}P^I_L(w)P^I_L(z)-{D \over (w-z)^2},$$
$$T_R(\bar z)=\lim_{\bar w\to \bar z}{1 \over 2}P^I_R(\bar w)P^I_R(\bar
z)-{D \over (\bar w-\bar z)^2},\eqno (2-1)$$
where $P^I_L(z)$ and $P^I_R(\bar z)$ are the momentum operators of the
left- and right-movers defined by
$$\eqalign{
P^I_L(z)&=i\partial_{z}X^I_L(z),\cr
P^I_R(\bar z)&=i\partial_{\bar z}X^I_R(\bar z), \quad (I=1,\dots,D).\cr}
\eqno(2-2)$$
It follows that the energy-momentum tensors are invariant under the action
of $g$ if $g$ acts on $(P^I_L(z),P^I_R(\bar z))$ as
$$g(P^I_L(z),P^I_R(\bar z))g^{\dagger}=(U^{IJ}_LP^J_L(z),U^{IJ}_RP^J_R(
\bar z)),\eqno(2-3)$$
where $U_L$ and $U_R$ are suitable elements of the $D$-dimensional
orthogonal group $O(D)$. Note that $U_L$ is not necessarily equal to $U_R$
and that orbifolds with $U_L\neq U_R$ are called asymmetric orbifolds
[13].

In the untwisted sector, the left- and right-moving string coordinates,
$X^I_L(z)$ and $X^I_R(\bar z)$, are expanded as
$$\eqalign{
X^I_L(z)&=x^I_L-ip^I_L lnz+i\sum_{n\neq 0}{1\over n}\alpha^I_{Ln}z^{-n},
\cr
X^I_R(\bar z)&=x^I_R-ip^I_R ln\bar z +i\sum_{n\neq 0}{1\over n}
\alpha^I_{Rn}\bar z^{-n},\quad(I=1,\cdots,D),\cr}\eqno(2-4)$$
where $x^I_L$ and $p^I_L$ ($x^I_R$ and $p^I_R$) are the center of mass
coordinate and momentum of the left- (right-) mover, respectively. The
quantization conditions are given by
$$[x^I_L,p^J_L]=i\delta^{IJ}=[x^I_R,p^J_R],$$
$$[\alpha^I_{Lm},\alpha^J_{Ln}]=m\delta^{IJ} \delta_{m+n,0}=[
\alpha^I_{Rm},\alpha^J_{Rn}],$$
$$otherwise \ zeros.\eqno(2-5)$$
The toroidal compactification means that the momentum $(p^I_L,p^I_R)$ lies
on a $(D+D)$-dimensional lorentzian even self-dual lattice $\Gamma^{D,D}$
[14]. In terms of $(p^I_L,\alpha^I_{Ln})$ and $(p^I_R,\alpha^I_{Rn})$, eq.
(2-3) can be rewritten as
$$\eqalign{
g(p^I_L,\alpha^I_{Ln})g^{\dagger}&=U^{IJ}_L(p^J_L, \alpha^J_{Ln}),\cr
g(p^I_R,\alpha^I_{Rn})g^{\dagger}&=U^{IJ}_R(p^J_R,\alpha^J_{Rn}).\cr}\eqno
(2-6)$$
Since $(p^I_L,p^I_R)$ lies on the lattice $\Gamma^{D,D}$, the action of
$g$ on $(p^I_L,p^I_R)$ should be an automorphism of $\Gamma^{D,D}$, i.e.,
$$( U^{IJ}_Lp^J_L, U^{IJ}_Rp^J_R)\in \Gamma^{D,D} \quad{\rm for \ all}
\quad (p^I_L, p^I_R)\in \Gamma^{D,D}.\eqno(2-7)$$

Since $P^I_L(z)$ and $P^I_R(\bar z)$ do not include $x^I_L$ and $x^I_R$,
the relation (2-3) or (2-6) does not completely determine the action of
$g$ on $(x^I_L,x^I_R)$. In fact, the general action of $g$ on $
(x^I_L,x^I_R)$, which is compatible with the quantization conditions
(2-5), may be given by [20]
$$g(x^I_L,x^I_R)g^{\dagger}=(U^{IJ}_L(x^J_L+\pi{\partial \Phi(p_L, p_R)
\over \partial p^J_L}) ,\ U^{IJ}_R(x^J_R+\pi{\partial \Phi(p_L,p_R) \over
\partial p^J_R})),\eqno(2-8)$$
where $\Phi(p_L,p_R)$ is an arbitrary function of $p^I_L$ and $p^I_R$. Let
$g_U$ be the unitary operator which satisfies
$$g_U(X^I_L(z),X^I_R(\bar z))g_U^{\dagger}=(U^{IJ}_LX^J_L(z),U^{IJ}_RX^J_R
(\bar z)), \eqno(2-9)$$
and
$$g_U|0> =|0>,\eqno(2-10) $$
where $|0>$ is the vacuum of the untwisted sector. Then, the twist
operator $g$ which generates the transformations (2-6) and (2-8) will be
given by
$$g=e^{i\pi\Phi(p_L ,p_R )}g_U.\eqno(2-11)$$
At this stage, $\Phi(p_L,p_R)$ is an arbitrary function of $p^I_L$ and
$p^I_R$. In section 4, we will see that the second condition (ii) severely
restricts the form of the phase factor in $g$.

It may be worth while making a comment on the path integral formalism
[21] here. The action of a closed bosonic string theory will be given by
\footnote{$^\dagger$}{$\eta^{\alpha \beta}=diag(1,-1)$ and $
\varepsilon^{01}=-\varepsilon^{10}=1$.}
$$S[X]=\int d\tau \int_{0}^{\pi} d\sigma {1\over2\pi}\{\eta^{\alpha \beta}
\partial _{\alpha} X^I \partial_{\beta}X^I+\varepsilon^{\alpha
\beta}B^{IJ}\partial_{\alpha}X^I \partial_{\beta}X^J\}, \eqno(2-12) $$
where $\tau$ and $\sigma$ correspond to the ``time'' and ``space''
variables of the world sheet and $B^{IJ}$ $(I,J=1,\cdots,D)$ is  an
antisymmetric constant background field [14]. The string coordinate $X^I(
\tau,\sigma)$ in the untwisted sector will be expanded as
$$ X^I(\tau,\sigma)=x^I+(p^I-B^{IJ}w^J)\tau+w^I\sigma + (oscillators),
\eqno(2-13)$$
where $p^I$ and $w^I$ are the center of mass momentum and the winding
number, which are related to the left- and the right-moving momenta,
$p^I_L$ and $p^I_R$, as follows:
$$p^I_L={1\over2}p^I+{1\over2}(1-B)^{IJ}w^J,$$
$$p^I_R={1\over2}p^I-{1\over2}(1+B)^{IJ}w^J.\eqno(2-14)$$
Let us consider a transformation
$$X^I \rightarrow U^{IJ}X^J,\eqno(2-15)$$
where $U^{IJ}\in O(D)$. (This corresponds to a symmetric orbifold, i.e.,
$U_L=U_R\equiv U$.) Clearly the action (2-12) is not invariant under the
transformation (2-15) unless
$$[B,U]=0. \eqno(2-16)$$
The noncommutativity of $B^{IJ}$ and $U^{IJ}$ might cause a trouble in the
path integral formalism because the action (2-12) will not be
single-valued on the orbifold. On the other hand, this noncommutativity
seems to cause no trouble in the operator formalism because the second
term in the action (2-12) is a total divergence and hence the explicit
$B^{IJ}$-dependence does not appear in the energy-momentum tensors (2-1)
as well as the equation of motion. The $B^{IJ}$-dependence can, however,
appear in the zero modes as in eqs.(2-13) and (2-14). As we will see in
section 4, the noncommutativity of $B^{IJ}$ and $U^{IJ}$ (more generally
see eq.(4-6) for asymmetric orbifolds) might cause a trouble even in the
operator formalism, that is, the violation of the duality of amplitudes.
Its resolution will be our main concern in section 4. In this paper, we
will restrict our considerations only to the operator formalism.
We will leave the reinterpretation of our results from the point of view
of the path integral formalism for future work.

\vskip 10mm

\section{3. A Representation of Cocycle Operators}

In this section, we shall investigate cocycle properties of vertex
operators and give an explicit operator representation of cocycle
operators. Let us consider a vertex operator which describes the emission
of a state with  the momentum $(k^I_L, k^I_R)\in \Gamma^{D,D},$
$$V(k_L,k_R;z)=:e^{ik_L\cdot X_L(z)+ik_R\cdot X_R(\bar z)}C_{k_L,k_R}:,
\eqno(3-1)$$
where : : denotes the normal ordering and $C_{k_L,k_R}$ is the cocycle
operator, which is attached to the vertex operator to ensure the correct
commutation relations and the duality of amplitudes [16,22]. The product
of two vertex operators
$$V(k_L,k_R;z)V(k'_L, k'_R;z'),\eqno(3-2)$$
is well-defined if $|z| > |z'|$. The different ordering of the two vertex
operators corresponds to the different ``time"-ordering. To obtain
scattering amplitudes, we must sum over all possible ``time"-ordering for
the emission of states. We must then establish that each contribution is
independent of the order of the vertex operators to enlarge the regions of
integrations over $z$ variables [15]. Thus the product (3-2), with respect
to $z$ and $z'$, has to be analytically continued to the region $|z'| >
|z| $ and to be identical to
$$V(k'_L, k'_R ; z')V(k_L ,k_R ;z ), \eqno(3-3)$$
for $ |z'| > |z| $.  In terms of the zero modes, the above statement can
be expressed as
$$ V_0(k_L,k_R)V_0(k'_L,k'_R)=(-1)^{k_L\cdot k'_L-k_R\cdot k'_R}V_0
(k'_L,k'_R)V_0(k_L,k_R),\eqno(3-4)$$
where
$$V_0(k_L,k_R)=e^{ik_L\cdot x_L+ik_R\cdot x_R}C_{k_L,k_R}.\eqno(3-5)$$
The factor $(-1)^{k_L\cdot k'_L-k_R\cdot k'_R}$ in eq.(3-4) appears in
reversing the order of the nonzero modes of the vertex operators. This
annoying factor is the reason for the necessity of the cocycle operator
$C_{k_L,k_R}$.

The second condition (ii) is now replaced by the statement that the
duality relation (3-4) has to be preserved under the action of $g$. To
examine this condition, we need to know an explicit operator
representation of the cocycle operator $C_{k_L,k_R}$. For notational
simplicity, we may use the following notations: $k^A \equiv (k^I_L,
k^I_R)$, $x^A \equiv (x^I_L, x^I_R),\dots$ etc. $(A, B, \dots$ run from 1
to $2D$ and $I, J,\dots$ run from 1 to $D$.) To obtain an operator
representation of the cocycle operator $C_k$, let us assume [23,24]
$$C_k=e^{i\pi k^A M^{AB}{\hat p}^B} ,\eqno(3-6)$$
where the wedge $\wedge$ may be attached to operators to distinguish
between c-numbers and q-numbers. Then, the matrix $M^{AB}$ has to satisfy
$$e^{i\pi k^A (M-M^T)^{AB}k'^B}=(-1)^{k^A \eta^{AB} k'^B}\qquad{\rm for \
all}\quad  k^A, k'^A \in  \Gamma ^{D,D},\eqno(3-7)$$
where
$$\eta^{AB}=\pmatrix{{\bf 1}&{\bf 0}\cr {\bf 0}& -{\bf 1}\cr}^{AB}.\eqno
(3-8)$$
A solution to this equation may be given by
$$M^{AB}={\pmatrix{-{1\over 2}B^{IJ}& -{1\over 2}(1-B)^{IJ} \cr {1\over 2}
(1+B)^{IJ} &-{1\over 2}B^{IJ} }}^{AB} ,\eqno(3-9)$$
which satisfies
$$M^{AB}=-M^{BA}.\eqno(3-10)$$
The $B^{IJ}$ is an antisymmetric constant matrix and is defined as
follows: Any $(D+D)$-dimensional lorentzian even self-dual lattice $
\Gamma^{D,D}$ can be parametrized in terms of a $D$-dimensional Euclidean
lattice $\Lambda$ and an antisymmetric constant matrix $B^{IJ}$ [14] as
\footnote{$^\dagger$}{The variables $p^I$, $w^I$ and $B^{IJ}$ are exactly
the same as in eq.(2-14).}
$$p^I_L={1\over2}p^I+{1\over2}(1-B)^{IJ}w^J,$$
$$p^I_R={1\over2}p^I-{1\over2}(1+B)^{IJ}w^J,\eqno(3-11)$$
where
$$\eqalignno{
(p^I_L,p^I_R)&\in\Gamma^{D,D},&\cr
p^I&\in 2\Lambda^*,&\cr
w^I&\in \Lambda.&(3-12)\cr}$$
Here, $\Lambda^*$ denotes the dual lattice of $\Lambda$. Physically, $p^I$
and $w^I$ correspond to the center of mass momentum and the winding
number, respectively.

Although we have obtained a representation of the cocycle operator $C_k$,
its representation is not unique. In fact, there exist infinitely many
other representations of $C_k$. However, as we will see in appendix B, by
a suitable unitary transformation any representation of $C_k$ can be shown
to reduce to eq.(3-6) with (3-9) up to a constant phase. Thus, it will be
sufficient to consider only the representation (3-6) with (3-9) for our
purpose. In the next section, we will see that the representation (3-6)
plays a crucial role in investigating the duality of amplitudes.

\vskip 10mm

\section{4. The Duality of Amplitudes}

In the previous section, we have obtained a representation of the cocycle
operator $C_k$. To explicitly show the dependence of the cocycle operator
in the zero mode part of the vertex operator (3-5), we may write
$$ V_0 (k ; M ) \equiv e^{ik\cdot \hat x}e^{i\pi k\cdot M \hat p}.\eqno
(4-1)$$
Under the action of $g_U$, $V_0 (k;M)$ transforms as
$$ g_UV_0 (k;M)g^{\dagger}_U =V_0 (U^T k; U^T M U),\eqno(4-2)$$
where
$$U^{AB}=\pmatrix{U^{IJ}_L & 0\cr 0 & U^{IJ} _R \cr}^{AB} .\eqno(4-3)$$
It is easy to see that the product of $V_0(k;M)$ and $V_0(k';U^TMU)$
satisfies
$$V_0 (k;M)V_0 (k';U^T M U )=\xi (-1)^{k\cdot \eta k'}V_0 (k';U^T M U )V_0
(k;M),\eqno(4-4)$$
where
$$\xi =e^{-i\pi k \cdot (M-U^T M U)k'}. \eqno(4-5)$$
This relation implies that the duality relation (3-4) cannot be preserved
under the action of $g_U $ unless $\xi =1$ for all $k^A,k'^A\in
\Gamma^{D,D}$. It may be worth while noting that if $\xi\ne1$ it means
$$[M,U]\ne0.\eqno(4-6)$$
For symmetric orbifolds (i.e., $U_L=U_R$), eq.(4-6) means the
noncommutativity of $B^{IJ}$ and $U^{IJ}_L$ (or $U^{IJ}_R$). As mentioned
in section 2, this noncommutativity may cause a trouble in the path
integral formalism and, as just seen above, also in the operator formalism
it causes a trouble, that is, the violation of the duality relation (3-4)
under the action of $g_U$.

We have seen that the duality relation (3-4) cannot be preserved under
the action of $g_U$ unless $\xi=1$ for all $k^A,k'^A\in\Gamma^{D,D}$. It
does not, however, mean the violation of the duality relation under the
action of $g$ because the freedom of $\Phi(p)$ in $g$ has not been used
yet. Define
$$\eqalignno{
V'_0(k;M)&\equiv gV_0(Uk;M)g^\dagger&\cr
&=e^{ik\cdot\hat x}e^{i\pi k\cdot U^TMU\hat p}e^{i\pi\Phi(\hat p+k)-i\pi
\Phi(\hat p)}.&(4-7)\cr}$$
It is easy to see that
$$V_0(k;M)V'_0(k';M)=e^{-i\pi\Theta}(-1)^{k\cdot\eta k'}V'_0(k';M)V_0
(k;M),\eqno(4-8)$$
where
$$\Theta=k^A(M-U^TMU)^{AB}k'^B+\Phi(p-k-k')-\Phi(p-k)-\Phi(p-k')+\Phi(p).
\eqno(4-9)$$
Thus the duality relation (3-4) requires that
$$\Theta=0 \quad {\rm mod} \ 2.\eqno(4-10)$$
To solve the equation (4-10), it may be convenient to change the basis of
the momentum $p^A\in\Gamma^{D,D}$. Let $e^A_a$ $(a=1,\cdots,2D)$ be a
basis of $\Gamma^{D,D}$, i.e.,
$$\Gamma^{D,D}=\{p^A=\sum_{a=1}^{2D}p^ae^A_a, \ p^a\in{\bf Z} \}.\eqno
(4-11)$$
Suppose that $\Phi(p)$ is expanded as
$$\Phi(p)=\Phi_0(p)+\Delta\Phi(p),\eqno(4-12)$$
where
$$\Phi_0(p)=\phi+2v_ap^a+{1\over2}C_{ab}p^ap^b,\eqno(4-13)$$
$$\Delta\Phi(p)=\sum_{n=2}^N{1\over n!}\Delta C_{a_1\cdots a_n}^{
(n)}p^{a_1}\cdots p^{a_n},\quad(p^a\in{\bf Z}).\eqno(4-14)$$
Here, $N$ $(\ge 2)$ is an arbitrary positive integer and the symmetric
matrix $C_{ab}$ is defined through the relation,
$$C_{ab}=-e^A_a (M-U^T M U )^{AB} e^B_b \quad{\rm mod}\  2 . \eqno(4-15)$$
At first sight, it seems that there is no solution to eq.(4-15) because
$C_{ab}$ is a symmetric matrix but $M^{AB}$ is an antisymmetric one.
However, we can always find a symmetric matrix $C_{ab}$ satisfying (4-15)
because eq.(3-7) with eq.(3-10) implies that
$$e^A_a(M-U^TMU)^{AB}e^B_b\in{\bf Z},\eqno(4-16)$$
which guarantees the existence of a solution to eq.(4-15). Inserting eq.
(4-12) into eq.(4-9) and using the relation (4-15), we find that the
condition (4-10) reduces to
$$\Delta\Phi(p-k-k')-\Delta\Phi(p-k)-\Delta\Phi(p-k')+\Delta\Phi(p)=0
\quad{\rm mod} \ 2.\eqno(4-17)$$

Inserting eq.(4-14) into eq.(4-17) and comparing the $N$th order terms of
both sides of eq.(4-17) with respect to $p^a$, $k^a$ and $k'^a$, we have
$${1\over N!}\sum_{a_1,\cdots,a_N=1}^{2D}\Delta C_{a_1\cdots a_N}^{(N)}\{
(p-k-k')^{a_1}\cdots(p-k-k')^{a_N}-(p-k)^{a_1}\cdots(p-k)^{a_N}$$
$$-(p-k')^{a_1}\cdots(p-k')^{a_N}+p^{a_1}\cdots p^{a_N} \} = 0 \quad {\rm
mod} \ 2 , \eqno(4-18)$$
for all $p^a,k^a,k'^a\in{\bf Z}$. This equation gives various constraints
on the coefficient $\Delta C_{a_1\cdots a_N}^{(N)}$. For example,
$$\Delta C_{a\cdots ab}^{(N)}\in (N-1)!\ 2{\bf Z},$$
$$\Delta C_{a\cdots abb}^{(N)}\in (N-2)!\ 2!\ 2{\bf Z},$$
$$\Delta C_{a\cdots abc}^{(N)}\in (N-2)!\ 2{\bf Z},$$
$$\Delta C_{a\cdots abbb}^{(N)}\in (N-3)!\ 3!\ 2{\bf Z},$$
$$ \cdots $$
$$ {\rm etc}.\eqno(4-19)$$
Then it is not difficult to show the following equality:
$${1\over N!}\sum_{a_1,\cdots,a_N=1}^{2D}\Delta C_{a_1\cdots a_N}^{(N)}
p^{a_1}\cdots p^{a_N} = {1\over N!}\sum_{a=1}^{2D}\Delta C_{a\cdots a}^{
(N)} (p^a)^N \quad {\rm mod} \ 2, \eqno(4-20)$$
with $\Delta C_{a\cdots a}^{(N)}\in (N-1)!\ 2{\bf Z}$. Let $(m,n)$ be the
highest common divisor of $m$ and $n$ and $\varphi(n)$ be the Euler
function, which is equal to the number of $d$ such that $(d,n)=1$ for $d
=1,2,\cdots,n-1$. In appendix A, we will prove the following theorem: Let
$n$ and $p$ be arbitrary positive integers. Then
$$p^n=p^{n-\varphi(n)} \quad {\rm mod} \ n.\eqno(4-21)$$
Using this theorem and noting $\Delta C_{a\cdots a}^{(N)}\in (N-1)!\ 2{\bf
Z}$, we have
$${1\over N!}\sum_{a=1}^{2D}\Delta C_{a\cdots a}^{(N)} (p^a)^N = {1\over
N!}\sum_{a=1}^{2D}\Delta C_{a\cdots a}^{(N)} (p^a)^{N-\varphi(N)} \quad {
\rm mod} \ 2. \eqno(4-22)$$
Thus we have found that the $N$th order term of $\Phi(p)$ (i.e., the left
hand side of eq.(4-20)) can reduce to the $(N-\varphi(N))$th order term
(i.e., the right hand side of eq.(4-22)). Therefore, we can put the $N$th
order term of $\Phi(p)$ to be equal to zero because the right hand side of
eq.(4-22) can be absorbed into the $(N-\varphi(N))$th order term of $\Phi
(p)$ by suitably redefining the coefficient of the $(N-\varphi(N))$th
order term of $\Phi(p)$.

Next consider the $(N-1)$th order term in eq.(4-17) with respect to
$p^a$, $k^a$ and $k'^a$. Comparing both sides of eq.(4-17) and using the
theorem (4-21), we can show the following equality:
$${1\over (N-1)!}\sum_{a_1,\cdots,a_{N-1}=1}^{2D}\Delta C_{a_1\cdots
a_{N-1}}^{(N-1)} p^{a_1}\cdots p^{a_{N-1}}\qquad\qquad\qquad\qquad\qquad$$
$$\qquad\qquad\qquad\qquad = {1\over (N-1)!}\sum_{a=1}^{2D}\Delta C_{a
\cdots a}^{(N-1)} (p^a)^{N-1-\varphi(N-1)} \quad {\rm mod} \ 2, \eqno
(4-23)$$
with $\Delta C_{a\cdots a}^{(N-1)}\in (N-2)!\ 2{\bf Z}$. Thus the $
(N-1)$th order term of $\Phi(p)$ can be absorbed into the $(N-1-\varphi
(N-1))$th order term of $\Phi(p)$ by suitably redefining the coefficient
of the $(N-1-\varphi(N-1))$th order term. Repeating the above argument
order by order, we conclude that $\Delta\Phi(p)$ can be put to be equal to
zero, i.e.,
$$\Delta\Phi(p)=0\quad{\rm mod} \ 2.\eqno(4-24)$$

We have observed that the duality relation can be preserved under the
action of $g$ if $\Phi(p)$ in $g$ is chosen as
$$\Phi(p)=\phi+2v_ap^a+{1\over2}p^aC_{ab}p^b,$$
or equivalently,
$$\Phi(p)=\phi+2v^A\eta^{AB}p^B+{1\over2}p^AC^{AB}p^B,\eqno(4-25)$$
where the symmetric matrix $C_{ab}$ is defined through the relation (4-15)
and
$$v_a=v^A\eta^{AB}e_a^B,$$
$$C_{ab}=e_a^AC^{AB}e_b^B.\eqno(4-26)$$
We will see in the next section that modular invariance requires $\phi=0$
and imposes some constraints on $v_a$. The symmetric matrix $C_{ab}$ seems
not to be defined uniquely in eq.(4-15). Let $C'_{ab}$ be another choice
satisfying eq.(4-15). Then, we find
$$\eqalignno{
{1 \over 2}\sum_{a,b}(C'_{ab}-C_{ab})p^ap^b&={1\over 2}\sum_{a=b}
(C'_{aa}-C_{aa})(p^a)^2\quad{\rm mod} \ 2 &\cr
& = {1\over 2} \sum_a (C'_{aa}-C_{aa})p^a \qquad{\rm mod} \ 2,&(4-27)
\cr}$$
where we have used the fact that $C'_{ab}-C_{ab}\in 2{\bf Z}$ and $p^a\in{
\bf Z}$. Thus the difference between $C'_{ab}$ and $C_{ab}$ can be
absorbed into the redefinition of $v_a$ and hence the choice of $C_{ab}$
is essentially unique. Therefore, it is concluded that any twist operator
$g$ can always be parametrized by $(U_L,v_L;U_R,v_R)$ and that the action
(1-3) of $g$ on the string coordinate $X^A=(X^I_L,X^I_R)$ in the untwisted
sector is not in general correct but
$$gX^Ag^\dagger=U^{AB}(X^B+2\pi\eta^{BC}v^C+\pi C^{BC}p^C),\eqno(4-28)$$
as announced in the introduction.

Does the third term of $\Phi(p)$ in eq.(4-25) affect the physical
spectrum? Any physical state on the orbifold has to be invariant under the
action of $g$. The third term in eq.(4-25) contributes to $g$ as a
momentum-dependent phase and hence plays an important role in extracting
physical states from the Hilbert space. Although we have introduced the
third term of $\Phi(p)$ in eq.(4-25) to preserve the duality of
amplitudes, we will see in the next section that modular invariance will
also require the introduction of the third term of $\Phi(p)$ in eq.(4-25).

\vskip 10mm

\section{5. One Loop Modular Invariance}

In this section, we will investigate one loop modular invariance of
partition functions. Let $Z(h,g;\tau)$ be the partition function of the
$h$-sector twisted by $g$ which is defined, in the operator formalism, by
$$Z(h,g;\tau)={\rm Tr}[g e^{i2\pi\tau(L_0-{D\over24})-i2\pi\bar\tau({\bar
L}_0-{D\over24})}]_{h-sector} , \eqno(5-1)$$
where $L_0 (\bar L_0 )$ is the Virasoro zero mode operator of the left-
(right-) mover. The trace in eq.(5-1) is taken over the Hilbert space of
the $h$-sector. Then, the one loop partition function will be of the form,
$$Z(\tau)={1\over N} \sum _{g,h\in G \atop gh=hg}Z(h,g;\tau) , \eqno
(5-2)$$
where $N$ is the order of $G$. In the above summation, only the elements
$h$ and $g$ which commute each other contribute to the partition function.
This will be explained as follows: To calculate $Z(h,g;\tau)$ in the
operator formalism, we need to introduce the string coordinate $(X^I_L
(z),X^I_R(\bar z))$ in the $h$-sector, which obeys the boundary condition
$$(X^I_L(e^{2\pi i }z) , X^I_R(e^{-2\pi i}\bar z)) = h\cdot (X^I_L (z),
X^I_R (\bar z)),\eqno (5-3)$$
up to torus shifts. Let us consider the action of $g$ on the string
coordinate in the $h$-sector. Then it turns out that $g(X^I_L(z),X^I_R(
\bar z))g^\dagger$ obeys the boundary condition of the $ghg^{-1}$-sector.
Let $|h>$ be any state in the $h$-sector. The above observation implies
that the state $g|h>$ belongs to the $ghg^{-1}$-sector but not the
$h$-sector (unless $g$ commutes with $h$). Therefore, in the trace formula
(5-1), $Z(h,g;\tau)$ will vanish identically unless $g$ commutes with $h$.

One loop modular invariance of the partition function is satisfied
provided
$$Z(h,g;\tau +1) = Z(h,hg;\tau ), \eqno(5-4)$$
$$Z(h,g;-{1\over \tau}) = Z(g^{-1},h;\tau ). \eqno(5-5)$$
Let us first consider the partition function of the untwisted sector
twisted by $g$, i.e., $Z(1,g;\tau)$. It follows from the discussions of
section 4 that in the untwisted sector the twist operator $g$ would be of
the form
$$g=e^{i\pi \Phi (p)}g_U  , \eqno(5-6)$$
where
$$\Phi(p)=\phi+2v^A\eta^{AB}p^B+{1\over2}p^AC^{AB}p^B.\eqno(5-7)$$
The symmetric matrix $C^{AB}$ is defined through the relation (4-15) or
$$k^A C^{AB} k'^B = -k^A (M-U^T M U)^{AB} k'^B \quad{\rm mod} \ 2,\eqno
(5-8)$$
for $k^A,k'^A\in\Gamma^{D,D}$. Let $n$ be the smallest positive integer
such that $g^n=1$. Then, it follows that
$$U^n ={\bf 1} , \eqno(5-9)$$
$$n\phi+\sum^{n-1}_{\ell =0}\lbrace 2v^A(\eta U^{\ell})^{AB}p^B + {1\over
2}p^A(U^{-{\ell}} C U^{\ell})^{AB}p^B \rbrace = 0 \quad{\rm mod} \ 2,\eqno
(5-10)$$
for all $p^A \in \Gamma ^{D,D}$. The zero mode part of $Z(1,g;\tau)$ can
easily be evaluated and the result is
$$Z(1,g;\tau)_{zero\  mode} = \sum _{(k_R, k_R)\in \Gamma ^{d,\bar d}_g}
e^{i\pi \Phi (k)} e^{i\pi \tau k^2 _L -i\pi \bar \tau k^2 _R}, \eqno
(5-11)$$
where $\Gamma_g^{d,\bar d}$ is the $g$-invariant sublattice of $
\Gamma^{D,D}$, i.e.,
$$\Gamma^{d,\bar d}_g = \{ (k_L^I,k_R^I) \in \Gamma^{D,D} |
(U_L^{IJ}k_L^J, U_R^{IJ}k_R^J) = (k_L^I , k_R^I)\} . \eqno(5-12)$$
Here, $(d, \bar d)$ denotes singature of the lorentzian lattice $
\Gamma^{d, \bar d}_g$ . We now show that the following relation holds for
a suitable constant vector $v'^A $ :
$$ {1\over 2 } k^A C^{AB} k^B = 2 v'^A \eta ^{AB} k^B \quad{\rm mod} \ 2 ,
\eqno(5-13)$$
for all $k^A \in \Gamma^{d,\bar d}_g $. To show this , define
$$f(k) \equiv {1\over 2}k^A C^{AB} k^B .\eqno(5-14)$$
Note that
$$\eqalignno{
k^A C^{AB} k'^B & = -k^A (M-U^T M U)^{AB} k'^B \quad{\rm mod} \ 2 &\cr
& = 0 \ {\rm mod} \ 2\quad{\rm for\  all} \ k^A , k'^A \in \Gamma ^{d,\bar
d }_g ,&(5-15)\cr}$$
where we have used eqs. (5-8) and (5-12).
It follows that
$$f(k +k') = f(k) + f(k') \quad{\rm mod} \ 2 , \eqno(5-16)$$
for all $k , k' \in \Gamma ^{d,\bar d }_g $. This relation ensures the
existence of a vector $v'$ satisfying eq.(5-13). Using the relation
(5-13), we can write eq.(5-11) as
$$Z(1,g;\tau)_{zero\  mode}=\sum_{(k_L,k_R)\in\Gamma^{d,\bar d}_g}e^{i\pi
\phi + i2\pi (v+v')\cdot \eta k} e^{i\pi \tau k^2_L - i\pi \bar \tau
k^2_R}. \eqno(5-17)$$
It will be useful to introduce a projection matrix ${\cal P}_U$ defined by
$${\cal P}_U = {1\over n } \sum^{n-1}_{\ell =0} U^{\ell}. \eqno(5-18)$$
Noting that ${\cal P}_U k = k $ for all $ k \in \Gamma ^{d,\bar d }_g $
and using the Poisson resummation formula, we have
$$\eqalignno{& Z(1,g;-{1\over\tau})_{zero\  mode } &\cr
& =e^{i\pi \phi } {{(-i \tau )^{d\over 2} ({i\bar\tau } )^{\bar d \over 2}
}\over V_{\Gamma ^{d,\bar d}_g}} \sum _{(q_L , q_R) \in \Gamma ^{d,{\bar
d}^*}_g - v^* - v'^*} e^{i\pi \tau q^2_L - i\pi \bar \tau q^2_R },&(5-19)
\cr}$$
where $v^* + v'^* \equiv {\cal P}_U (v + v')$, $V_{\Gamma}$ denotes the
unit volume of the lattice $\Gamma $ and $\Gamma^{d,{\bar d}^*}_g $ is the
dual lattice of $\Gamma^{d,{\bar d}}_g $. It follows from eq.(5-19) that
we can easily extract information about the zero modes of the
$g^{-1}$-sector because $Z(1,g;\tau)$ should be related to $Z(g^{-1},1;
\tau)$ through the modular transformation, i.e.,
$$Z(g^{-1} , 1 ; \tau )= Z(1 , g ; -{1 \over \tau }) . \eqno (5-20)$$
The degeneracy of the ground state in the $g^{-1}$-sector may be given by
[13]
$${\sqrt{det'(1-U)}\over V_{\Gamma ^{d,\bar d}_g } } , \eqno(5-21)$$
where the determinant should be taken over the nonzero eigenvalues of
$1-U$ and the factor $\sqrt{det'(1-U)}$ will come from the oscillators.
The eigenvalues of the momentum $(q_L, q_R)$ in the $g^{-1}$-sector may be
given by
$$(q_L , q_R ) \in \Gamma^{d,\bar d*}_g - v^* - v'^* . \eqno(5-22)$$
It should be noted that the momentum eigenvalues in the $g^{-1}$-sector
are not given by $\Gamma_g^{d,\bar d *}-v^*$, which might naively be
expected [13]. The origin of the extra contribution $-v'^*$ is the third
term in eq.(5-7), which has been introduced to ensure the duality relation
of vertex operators.  As we will see later, this extra contribution to the
momentum eigenvalues becomes important in the left-right level matching
condition.

Information about the zero modes given above is sufficient to obtain $Z
(g^{-1},1;\tau)$ because the oscillator part of $Z(g^{-1},1;\tau)$ can
unambiguously be calculated. Then, it turns out that the relation (5-20)
puts a constraint on $\phi $ in eq. (5-7), i.e.,
$$\phi =0 . \eqno (5-23)$$
This is desirable because otherwise the vacuum in the untwisted sector
would not be invariant under the action of $g$ and hence would be removed
from the physical Hilbert space. In the point of view of the conformal
field theory, the vacuum in the untwisted sector will correspond to the
identity operator, which should be included in the operator algebra.

A necessary condition for modular invariance is the left-right level
matching condition [13,25]
$$Z(g^{-1},h;\tau+n)=Z(g^{-1},h;\tau).\eqno(5-24),$$
where $n$ is the smallest positive integer such that $g^n=1$. It follows
from eq.(5-1) that the level matching condition is satisfied only if
$$2n(L_0-\bar L_0)=0 \quad {\rm mod} \ 2,\eqno(5-25)$$
where $L_0$ $(\bar L_0)$ is the Virasoro zero mode operator of the left-
(right-) mover in the $g^{-1}$-sector. Since any contribution to $L_0$ and
$\bar L_0$ from the oscillators is a fraction of $n$, the level matching
condition can be written as
$$2n(\varepsilon_{g^{-1}}-\bar\varepsilon_{g^{-1}}+{1\over2}q_L^2-{1
\over2}q_R^2)=0 \quad {\rm mod} \ 2, \qquad {\rm for \ all} \ (q_L,q_R)\in
\Gamma^{d,\bar d^*}_g-v^*-{v'}^*,\eqno(5-26)$$
where $(\varepsilon_{g^{-1}},\bar\varepsilon_{g^{-1}})$ is the conformal
dimension (or the zero point energy) of the ground state in the
$g^{-1}$-sector and is explicitly given by [1]
$$\varepsilon_{g^{-1}}={1\over4}\sum_{a=1}^D\rho_a(1-\rho_a),$$
$$\bar\varepsilon_{g^{-1}}={1\over4}\sum_{a=1}^D\bar\rho_a(1-\bar\rho_a).
\eqno(5-27)$$
Here, $exp(i2\pi\rho_a)$ and $exp(i2\pi\bar\rho_a)$ $(a=1,\cdots,D)$ are
the eigenvalues of $U_L$ and $U_R$ with $0\le\rho_a,\bar\rho_a<1$,
respectively.

The condition (5-26) can further be shown to reduce to
$$2n(\varepsilon_{g^{-1}}-\bar\varepsilon_{g^{-1}}+{1\over2}(v^*_L
+{v'}_L^*)^2-{1\over2}(v^*_R+{v'}_R^*)^2)=0 \quad {\rm mod} \ 2.\eqno
(5-28)$$
To see this, we first note that ${\Gamma^{d,\bar d}_g}^*$ can be expressed
as [13]
$$\eqalignno{
 {\Gamma^{d,\bar d}_g}^*&={\cal P}_U\Gamma^{D,D}&\cr
          &=\{q^A={\cal P}_Uk^A, \ k^A\in\Gamma^{D,D}\}.&(5-29)\cr}$$
This follows from the property that $\Gamma^{D,D}$ is self-dual. From eq.
(5-29), any momentum $q^A\in{\Gamma^{d,\bar d}_g}^*-v^*-{v'}^*$ can be
parametrized as
$$q^A={\cal P}_U(k-v-v')^A \quad {\rm for \ some }\ k^A\in\Gamma^{D,D}.
\eqno(5-30)$$
Then, we have
$$\eqalign{
&n(q_L^2-q_R^2)=nq^A\eta^{AB}q^B\cr
&=nk^A(\eta{\cal P}_U)^{AB}k^B-2n(v+v')^A(\eta{\cal P}_U)^{AB}k^B+n(v^*
+{v'}^*)^A\eta^{AB}(v^*+{v'}^*)^B,\cr}$$
$$\eqno(5-31)$$
where we have used the relations
$${\cal P}_U\eta=\eta{\cal P}_U,$$
$${\cal P}_U^2={\cal P}_U,$$
$${\cal P}_U^T={\cal P}_U.\eqno(5-32)$$
Since $\Gamma^{D,D}$ is an even integral lattice and $U$ is an orthogonal
matrix satisfying $U^n={\bf 1}$, the first term in the right handed side
of eq.(5-31) is easily shown to reduce to
$$ nk^A(\eta{\cal P}_U)^{AB}k^B=\cases{k^A(\eta U^{n\over2})^{AB}k^B \quad
& mod 2  if  $n=\rm even$,\cr
0 \quad & mod  2  if $n=\rm odd$. \cr} \eqno(5-33)$$
Using the relation (5-13) and noting that $n{\cal P}_Uk\in\Gamma^{d,\bar
d}_g$, we can rewrite the second term in the right hand side of eq.(5-31)
as
$$-2n(v+v')^A(\eta{\cal P}_U)^{AB}k^B=-2nv^A(\eta{\cal P}_U)^{AB}k^B-{1
\over2}k^A\sum_{\ell=0}^{n-1}\sum_{m=0}^{n-1}(U^{-\ell}CU^m)^{AB}k^B \ {
\rm mod} \ 2.\eqno(5-34)$$
Replacing $p$ by $p+p'$ in eq.(5-10) with eq.(5-23) and then using eq.
(5-10) again, we have
$$p^A\sum_{\ell=0}^{n-1}(U^{-\ell}CU^\ell)^{AB}p'^B=0 \quad {\rm mod} \ 2,
\eqno(5-35)$$
for all $p,p'\in\Gamma^{D,D}$. For $n$ odd, it is not difficult to show
that
$$-2n(v+v')^A(\eta{\cal P}_U)^{AB}k^B=0 \quad {\rm mod} \ 2.\eqno(5-36)$$
To derive eq.(5-36), we will use eqs.(5-10), (5-23), (5-34) and (5-35).
For $n$ even, we will find
$$-2n(v+v')^A(\eta{\cal P}_U)^{AB}k^B=-k^A\sum_{\ell=0}^{{n\over2}-1}(U^{-
\ell}CU^{\ell+{n\over2}})^{AB}k^B \quad {\rm mod} \ 2.\eqno(5-37)$$
Remembering the relations (3-7), (3-10) and (5-8), we can finally find
that for $n$ even
$$-2n(v+v')^A(\eta{\cal P}_U)^{AB}k^B=k^A(\eta U^{n\over2})^{AB}k^B \quad
{\rm mod} \ 2.\eqno(5-38)$$
Combining \ the \ results \ (5-33), (5-36) \ and \ (5-38) \ and \ using \
the \ fact \ that $k^A(\eta U^{n\over2})^{AB}k^B\in{\bf Z}$, we have
$$nk^A(\eta{\cal P}_U)^{AB}k^B-2n(v+v')^A(\eta{\cal P}_U)^{AB}k^B=0 \quad
{\rm mod} \ 2.\eqno(5-39)$$
This completes the proof of (5-28).

We have shown that the left-right level matching condition (5-24) reduces
to the condition (5-28), which puts a restriction on the shift vector $v=
(v_L,v_R)$. It should be noticed that the level matching condition (5-28)
is not always satisfied for asymmetric orbifold models but trivially
satisfied for symmetric ones because $\varepsilon_{g^{-1}}=\bar
\varepsilon_{g^{-1}}$ and $(v_L^*+{v'}_L^*)^2=(v_R^*+{v'}_R^*)^2$ for
symmetric orbifold models. For the case of $C^{AB}=0$ in eq.(5-7), it has
been proved, in refs. [13,25], that the level matching condition is a
necessary and also sufficient condition for one loop modular invariance.
Even for the case of nonzero $C^{AB}$, the sufficiency can probably be
shown by arguments similar to refs. [13,25] although to this end we need
to know the action of $g$ on the string coordinate in every twisted
sector.

It should be emphasized that the third term in eq.(5-7) plays an
important role in the level matching condition because the relation (5-39)
might not hold in general if we put $v'$ to be zero, that is, $C_{ab}$ to
be zero by hand. In the next section, we will see such an example that the
introduction of the third term in eq.(5-7) makes partition functions
modular invariant.

\vskip 10mm

\section{6. An Example}

In this section, we shall investigate a symmetric ${\bf Z}_2$-orbifold
model in detail, which will give a good illustration of our formalism.
Many other examples can be found in ref.[26].

Let us introduce the root lattice $\Lambda_R$ and the weight lattice $
\Lambda_W$ of $SU(3)$ as
$$\Lambda_R=\{p^I=\sum_{i=1}^2 n^i\alpha^I_i, n^i \in {\bf Z}\},$$
$$\Lambda_W=\{p^I=\sum_{i=1}^2 m_i\mu^{iI}, m_i \in {\bf Z}\},\eqno(6-1)$$
where $\alpha_i$ and $\mu^i$ $(i=1,2)$ are a simple root and a fundamental
weight satisfying $\alpha_i\cdot\mu^j=\delta_i^j$. We will take $\alpha_i$
and $\mu^i$ to be
$$\alpha_1=({1\over\sqrt 2},\sqrt{3\over2}),$$
$$\alpha_2=({1\over\sqrt 2},-\sqrt{3\over2}),$$
$$\mu^1=({1\over\sqrt 2},\sqrt{1\over6}),$$
$$\mu^2=({1\over\sqrt 2},-\sqrt{1\over6}).\eqno(6-2)$$
The left- and right-moving momentum $(p^I_L,p^I_R)$ $(I=1,2)$ is defined
by eq.(3-11), i.e.,
$$p^I_L={1\over2}p^I+{1\over2}(1-B)^{IJ}w^{J},$$
$$p^I_R={1\over2}p^I-{1\over2}(1+B)^{IJ}w^{J},\eqno(6-3)$$
where $p^I$ and $w^I$ are the center of mass momentum and the winding
number, respectively and are assumed to lie on the following lattices:
$$p^I\in 2\Lambda_W,$$
$$w^I\in \Lambda_R.\eqno(6-4)$$
The antisymmetric constant matrix $B^{IJ}$ is chosen as
$$B^{IJ}=\pmatrix{0 & -{1\over\sqrt{3}} \cr
                  {1\over\sqrt{3}} & 0 \cr}.\eqno(6-5)$$
Then, it turns out that $(p^I_L,p^I_R)$ lies on the following $(2
+2)$-dimensional lorentzian even self-dual lattice $\Gamma^{2,2}$:
$$\Gamma^{2,2}=\{(p^I_L,p^I_R)|p^I_L,p^I_R\in \Lambda_W, p^I_L-p^I_R\in
\Lambda_R\}.\eqno(6-6)$$

We consider the following ${\bf Z}_2$-transformation:
$$g_U(X^I_L,X^I_R)g_U^\dagger=(U^{IJ}_LX^J_L,U^{IJ}_RX^J_R),\eqno(6-7)$$
where
$$U^{IJ}_L=U^{IJ}_R=\pmatrix{1 & 0 \cr
                             0  & -1 \cr }. \eqno(6-8)$$
This is an automorphism of $\Gamma^{2,2,}$, as it should be. According to
our prescription, the ${\bf Z}_2$-twist operator $g$ will be of the form
$$g=e^{i{\pi\over2}p^AC^{AB}p^{B}}g_U,\eqno(6-9)$$
where $p^A\equiv(p^I_L,p^I_R)$ and the symmetric matrix $C^{AB}$ is
defined through the relation
$$p^AC^{AB}p'^B=-p^A(M-U^TMU)^{AB}p'^B \quad {\rm mod} \ 2,\eqno(6-10)$$
for $p^A, p'^A \in \Gamma^{2,2}$. Here, we have taken a shift vector to
zero for simplicity and $M^{AB}$, $U^{AB}$ are defined by
$$M^{AB}=\pmatrix{-{1\over2}B^{IJ} & -{1\over2}(1-B)^{IJ} \cr
                  {1\over2}(1+B)^{IJ} & -{1\over2}B^{IJ} \cr}^{AB}, $$
$$U^{AB}=\pmatrix{U^{IJ}_L & 0 \cr
                  0  &  U^{IJ}_R  \cr}^{AB}.\eqno(6-11)$$
For symmetric orbifolds $(U_L=U_R)$, the defining relation (6-10) of
$C^{AB}$ may be replaced by
$$(p_L-p_R)^IC^{IJ}(p'_L-p'_R)^J={1\over2}(p_L-p_R)^I(B-U^T_LBU_L)^{IJ}
(p'_L-p'_R)^J \quad {\rm mod} \ 2, \eqno(6-12)$$
where $C^{AB}$ has been assumed to be of the form
$$C^{AB}=\pmatrix{C^{IJ} & -C^{IJ} \cr
                  -C^{IJ} & C^{IJ} \cr}^{AB}.\eqno(6-13)$$
Thus, the twist operator (6-9) can be written as
$$g=e^{i{\pi\over2}(p_L-p_R)^IC^{IJ}(p_L-p_R)^J}g_U.\eqno(6-14)$$
Since $p^I_L-p^I_R\in\Lambda_R$, the equation (6-12) may be rewritten as
$$\alpha^I_iC^{IJ}\alpha^J_j={1\over2}\alpha^I_i(B-U^T_LBU_L)^{IJ}
\alpha^J_j \quad {\rm mod} \ 2.\eqno(6-15)$$
The right hand side of eq.(6-15) is found to be
$${1\over2}\alpha^I_i(B-U^T_LBU_L)^{IJ}\alpha^J_j=\pmatrix{0 & 1 \cr
                                                          -1 & 0
\cr}_{ij},\eqno(6-16)$$
and hence $C^{IJ}$ cannot be chosen to be zero. We may choose $C^{IJ}$ as
$$\alpha^I_iC^{IJ}\alpha^J_j=\pmatrix{0 & 1 \cr
                                      1 & 0 \cr}_{ij},$$
or
$$C^{IJ}=\pmatrix{1 & 0 \cr
                  0 & -{1\over3} \cr}^{IJ}.\eqno(6-17)$$
This choice turns out to be consistent with $g^2=1$.

Let us consider the following momentum and vertex operators of the
left-mover:
$$P^I_L(z)=i\partial_zX^I_L(z),$$
$$V_L(\alpha ; z)=: e^{i\alpha\cdot X_L(z)}C_\alpha:,\eqno(6-18)$$
where $\alpha$ is a root vector of $SU(3)$ and $C_\alpha$ denotes a
cocycle operator. These operators form level one $SU(3)$ Ka{\v c}-Moody
algebra [22]. Under the action of $g$, they transform as
$$gP^I_L(z)g^\dagger=U^{IJ}_LP^J_L(z),$$
$$gV_L(\pm\alpha_1;z)g^\dagger=V_L(\pm\alpha_2;z),$$
$$gV_L(\pm\alpha_2;z)g^\dagger=V_L(\pm\alpha_1;z),$$
$$gV_L(\pm(\alpha_1+\alpha_2);z)g^\dagger=-V_L(\pm(\alpha_1+\alpha_2);z).
\eqno(6-19)$$
Thus, the ${\bf Z}_2$-invariant physical generators may be given by
$$\eqalign{
J_3(z)&=2P^1_L(z),\cr
J_\pm(z)&=\sqrt{2}(V_L(\pm\alpha_1;z)+V_L(\pm\alpha_2;z)).\cr}\eqno
(6-20)$$
These generators are found to form level four $SU(2)$ Ka{\v c}-Moody
algebra [27]. Note that the vertex operators $V_L(\pm(\alpha_1+
\alpha_2);z)$ are not invariant under the action of $g$ and hence they are
removed from the physical generators although the root vector $\alpha_1+
\alpha_2$ is invariant under the action of $g_U$.

We now examine one loop modular invariance of the partition function
which will be given by
$$Z(\tau)={1\over2}\sum_{\ell,m=0}^1Z(g^\ell,g^m;\tau),\eqno(6-21)$$
where
$$Z(g^\ell,g^m;\tau)={\rm Tr}[g^me^{i2\pi\tau(L_0-{2\over24})-i2\pi{\bar
\tau}(\bar L_0-{2\over24})}]_{g^\ell{\rm -sector}}.\eqno(6-22)$$
The partition functions of the untwisted sector can easily be evaluated
and the result is
$$Z(1,1;\tau)={1\over|\eta(\tau)|^4}\sum_{(k_L,k_R)\in\Gamma^{2,2}}e^{i\pi
\tau k^2_L-i\pi{\bar\tau}k^2_R},\eqno(6-23)$$
$$Z(1,g;\tau)={|\vartheta_3(0|\tau)\vartheta_4(0|\tau)|\over|\eta(
\tau)|^4}\sum_{(k_L,k_R)\in\Gamma^{1,1}_g}e^{i2\pi(v'_Lk_L-v'_Rk_R)}e^{i
\pi\tau k^2_L-i\pi{\bar\tau}k^2_R},\eqno(6-24)$$
where
$$v'_L=v'_R={1\over2\sqrt2},$$
$$\Gamma^{1,1}_g=\{(k_L,k_R)=(\sqrt2n,\sqrt2n'),n,n'\in{\bf Z}\}.\eqno
(6-25)$$
Here, the shift vector $(v'_L,v'_R)$ has been introduced through the
relation (5-13). The function $\eta(\tau)$ and $\vartheta_a(\nu|\tau)$ $(a
=1,\cdots,4)$ are the Dedekind $\eta$-function and the Jacobi theta
function:
$$\eta(\tau)=q^{1/12}\prod_{n=1}^\infty(1-q^{2n}), \qquad (q=e^{i\pi
\tau}),$$
$$\vartheta_{ab}(\nu|\tau)=\sum_{n=-\infty}^\infty exp\{i\pi(n+a)^2\tau+i2
\pi(n+a)(\nu+b)\},$$
$$\eqalign{
\vartheta_1(\nu|\tau)&=\vartheta_{{1\over2}{1\over2}}(\nu|\tau),\cr
\vartheta_2(\nu|\tau)&=\vartheta_{{1\over2}0}(\nu|\tau),\cr
\vartheta_3(\nu|\tau)&=\vartheta_{00}(\nu|\tau),\cr
\vartheta_4(\nu|\tau)&=\vartheta_{0{1\over2}}(\nu|\tau).\cr}\eqno(6-26)$$

It follows from the arguments given in section 5 that the degeneracy of
the ground state in the $g$-sector is
$${\sqrt{det'(1-U)}\over V_{\Gamma^{1,1}_g}}=1,\eqno(6-27)$$
and that the momentum eigenvalues in the $g$-sector are given by
$$(q_L,q_R)\in{\Gamma^{1,1}_g}^*-(v'_L,v'_R),\eqno(6-28)$$
where
$${\Gamma^{1,1}_g}^*=\{(q_L,q_R)=({1\over\sqrt2}n,{1\over\sqrt2}n'),n,n'
\in{\bf Z}\}.\eqno(6-29)$$
This information is sufficient to obtain $Z(g,1;\tau)$ and $Z(g,g;\tau)$:
$$Z(g,1;\tau)={|\vartheta_3(0|\tau)\vartheta_2(0|\tau)|\over2|\eta(
\tau)|^4}\sum_{(q_L,q_R)\in{\Gamma^{1,1}_g}^*-(v'_L,v'_R)}e^{i\pi\tau
q^2_L-i\pi{\bar\tau}q^2_R},\eqno(6-30)$$
$$Z(g,g;\tau)={|\vartheta_4(0|\tau)\vartheta_2(0|\tau)|\over2|\eta(
\tau)|^4}\sum_{(q_L,q_R)\in{\Gamma^{1,1}_g}^*-(v'_L,v'_R)}e^{i\pi
(q_L^2-q_R^2)}e^{i\pi\tau q^2_L-i\pi{\bar\tau}q^2_R}.\eqno(6-31)$$
It is easily verified that $Z(g^\ell,g^m;\tau)$ satisfies the following
desired relations:
$$Z(g^\ell,g^m;\tau+1)=Z(g^\ell,g^{m+\ell};\tau),$$
$$Z(g^\ell,g^m;-{1\over\tau})=Z(g^{-m},g^\ell;\tau),\eqno(6-32)$$
and hence the partition function (6-21) is modular invariant. It should be
emphasized that the existence of the shift vector $(v'_L,v'_R)$ ensures
modular invariance of the partition function: The level matching condition
$$Z(g,1;\tau+2)=Z(g,1;\tau),\eqno(6-33)$$
is satisfied because for all $(q_L,q_R)\in{\Gamma^{1,1}_g}^*-(v'_L,v'_R)$,
$$4({1\over2}q_L^2-{1\over2}q_R^2)=0 \quad {\rm mod } \ 2.\eqno(6-34)$$
If we put the shift vector $(v'_L,v'_R)$ or $C^{IJ}$ in $g$ to be zero by
hand, the level matching condition might, however, be destroyed because
eq.(6-34) dose not hold for $(q_L,q_R)\in{\Gamma^{1,1}_g}^*$.

It is interesting to note that in terms of the theta functions the
partition function obtained above can be expressed as
$$\eqalignno{
Z(1,1;\tau)&={1\over|\eta(\tau)|^4}\sum_{(k_L,k_R)\in\Gamma^{2,2}}e^{i\pi
\tau k^2_L-i\pi{\bar\tau}k^2_R},&\cr
Z(1,g;\tau)&={|\vartheta_3(0|\tau)\vartheta_4(0|\tau)|^2\over|\eta(
\tau)|^4},&\cr
Z(g,1;\tau)&={|\vartheta_3(0|\tau)\vartheta_2(0|\tau)|^2\over|\eta(
\tau)|^4},&\cr
Z(g,g;\tau)&={|\vartheta_4(0|\tau)\vartheta_2(0|\tau)|^2\over|\eta(
\tau)|^4}.&(6-35)\cr}$$
This partition function is exactly identical to that of a symmetric ${\bf
Z'_2}$-orbifold model whose ${\bf Z'_2}$-transformation is defined by
$${\bf Z'_2}:(X^I_L,X^I_R)\rightarrow(-X^I_L,-X^I_R),\eqno(6-36)$$
instead of the ${\bf Z}_2$-transformation (6-7). In this orbifold model,
level one $SU(3)$ Ka{\v c}-Moody algebra can be shown to ``break'' to
level four $SU(2)$ Ka{\v c}-Moody algebra. Thus, although the two orbifold
models are defined by the different ${\bf Z}_2$-transformations (6-7) and
(6-36), they give the same spectrum and interaction [28].

\vskip 10mm

\section{7. Discussions}

In this paper, we have investigated the following three consistency
conditions in detail: (i) the invariance of the energy-momentum tensors
under the action of the twist operators, (ii) the duality of amplitudes
and (iii) modular invariance of partition functions. From the analysis of
the second condition (ii), we have obtained various important results. The
following two points are probably main results of this paper: The first
point is the discovery of the third term in eq.(5-7), which is necessary
to preserve the duality of amplitudes under the action of $g$ and which
plays an important role in modular invariance of partition functions.
The second point is that any twist operator $g$ has been proved to be
represented by eq.(1-2). To show this, we have seen that the first
condition (i) is not sufficient and that the second condition (ii) is
crucial to restrict the allowed form of $\Phi(p_L,p_R)$ to eq.(5-7). It
should be emphasized that it is very important to show that by a suitable
unitary transformation any representation of cocycle operators can reduce
to the representation (3-6) with eq.(3-9) up to a constant phase because
our analysis has heavily relied on the representation (3-6).

We have found that the string coordinate $X^A=(X^I_L,X^I_R)$ in the
untwisted sector transforms under the action of $g$ as
$$gX^Ag^\dagger=U^{AB}(X^B+2\pi\eta^{BC}v^C+\pi C^{BC}p^C).\eqno(7-1)$$
It seems that the third term of eq.(7-1) has no clear geometrical meaning.
Although the momentum and vertex operators definitely transform under the
action of $g$, why does not the string coordinate transform definitely?
The reason is probably that in the point of view of the conformal field
theory the string coordinate is not a primary field and it is not a
well-defined variable on a torus. Thus,  there may be no reason why the
string coordinate itself should definitely transform under the action of
$g$. On the other hand, since the momentum and vertex operators are
primary fields and are well-defined on a torus, they should definitely
transform under the action of $g$. In fact, they transform as
$$ g(P^I_L(z),P^I_R(\bar z))g^\dagger = (U^{IJ}_LP^J_L(z),U^{IJ}_RP^J_R(
\bar z)),$$
$$ gV(k_L,k_R;z)g^\dagger = e^{i2\pi v\cdot\eta U^Tk+i{\pi\over 2}k\cdot
UCU^Tk}V(U_L^Tk_L,U_R^Tk_R;z). \eqno(7-2)$$
It may be worth while pointing out that the ``center of mass coordinate''
$x^A=(x^I_L,x^I_R)$ always appears as the following combination:
$$ x'^A \equiv x^A + \pi M^{AB}p^B, \eqno(7-3) $$
in the vertex operators and that $x'^A$ definitely transform under the
action of $g$, i.e.,
$$ gx'^Ag^\dagger = U^{AB}(x'^B+2\pi\eta^{BC}v^C), \eqno(7-4)$$
up to torus shifts although $x^A$ itself does not. This observation
strongly suggests that the variable $x'^A$ is more fundamental than $x^A$
[24,26].

We have succeeded to obtain the general class of bosonic orbifold models.
The generalization to superstring theories will be straightforward because
fermionic fields will definitely transform under the action of twist
operators.

We have restricted our considerations mainly to the untwisted sector.
However, much information about twisted sectors, in particular, zero
modes, can be obtained through modular transformations. Such information
is sufficient to obtain the partition function of the $g$-sector $Z(g,1;
\tau)$ but not $Z(g,h;\tau)$ in general because we have not constructed
twist operators in each twisted sector. The twist operator $g$ in the
$g$-sector can, however, be found to be of the form
$$g=e^{i2\pi(L_0-\bar L_0)}.\eqno(7-5)$$
This follows from the relation
$$Z(g,g;\tau)=Z(g,1;\tau+1).\eqno(7-6)$$
To obtain an explicit operator representation of any twist operator in
every twisted sector, we may need to construct vertex operators in every
twisted sector as in the untwisted sector. In the construction of vertex
operators in twisted sectors, the most subtle part is a realization of
cocycle operators. In the case of $\xi=1$ in eq.(4-5), (untwisted state
emission) vertex operators in any twisted sector have already been
constructed with correct cocycle operators in ref.[29]. In the case of $
\xi\ne1$, the prescription given in ref.[29] will be insufficient to
obtain desired vertex operators because the duality relation will not be
satisfied. Some attempts [30] have been made but the general construction
of correct vertex operators is still an open problem.

As mentioned in section 2, there might appear a trouble in the path
integral formalism unless eq.(2-16) is satisfied. Our success in the
operator formalism, however, probably means that our results can be
reinterpreted from the path integral point of view. Then the geometrical
meaning will become clear.

\endpage

\section{Appendix A}
In this appendix, we shall prove the following theorem: Let $n$ be a
positive integer. Then, for any positive integer $p$,
$$p^n=p^{n-\varphi(n)} \quad {\rm mod} \ n,\eqno(A-1)$$
where $\varphi(n)$ is the Euler function which is equal to the number of
$d$ such that $(d,n)=1$ for $d=1,2,\cdots,n-1$. Here, $(d,n)$ denotes the
highest common divisor of $d$ and $n$. The Euler function satisfies the
following relation:
$$\varphi(mn)=\varphi(m)\varphi(n) \quad {\rm if} \ (m,n)=1.\eqno(A-2)$$
To prove the theorem (A-1), we start with the Euler's theorem:
$$p^{\varphi(n)}=1 \quad {\rm mod} \ n \quad {\rm for} \ (p,n)=1.\eqno
(A-3)$$
Suppose that $n$ is decomposed as
$$n=(q_1)^{\ell_1}(q_2)^{\ell_2}\cdots(q_r)^{\ell_r},\eqno(A-4)$$
where $q_i$ $(i=1,\cdots,r)$ is a prime number and $q_i\ne q_j$ if $i\ne
j$. Let $\ell_{max}$ be the maximum number in the set of $\{\ell_i,i=1,
\cdots,r\}$. In terms of $q_i$, any positive integer $p$ can be decomposed
as
$$p=(q_1)^{\ell_1'}\cdots(q_r)^{\ell_r'}s,\eqno(A-5)$$
where $\ell_i'\ge0$ and $s$ is an integer such that $(s,n)=1$. Then it is
not difficult to show that
$$(q_i)^{\ell_i'(\varphi(n)+\ell_{max})}=(q_i)^{\ell_i'\ell_{max}} \quad {
\rm mod} \ n,$$
$$s^{\varphi(n)+\ell_{max}}=s^{\ell_{max}} \quad {\rm mod} \ n.\eqno
(A-6)$$
It follows that for any positive integer $p$
$$p^{\varphi(n)+\ell_{max}}=p^{\ell_{max}} \quad {\rm mod} \ n.\eqno
(A-7)$$
The Euler function $\varphi(n)$ satisfies
$$\sum_{1\le d\le n \atop d|n}\varphi(d)=n,\eqno(A-8)$$
where $d|n$ means that $d$ is a divisor of $n$. For any prime number
$q_i$,
$$\eqalignno{
\varphi(q_i^{\ell_i})&=q_i^{\ell_i-1}(q_i-1)&\cr
                     &\ge q_i^{\ell_i-1}&\cr
                     &\ge \ell_i.&(A-9)\cr}$$
Then, it follows from eqs.(A-8) and (A-9) that
$$n>\ell_{max}+\varphi(n).\eqno(A-10)$$
Multiplying (A-7) by $p^{n-\varphi(n)-\ell_{max}}$ and noting $n-\varphi
(n)-\ell_{max}>0$, we finally have
$$p^n=p^{n-\varphi(n)} \quad {\rm mod} \ n.\eqno(A-11)$$

\vskip 10mm

\section{Appendix B}
In this appendix, we shall prove that by a suitable unitary
transformation any representation of the cocycle operator $C_k$ can reduce
to
$$C_k=e^{i\pi k^AM^{AB}\hat p^B},\eqno(B-1)$$
with eq.(3-9) up to a constant phase.

We first note that the following factor:
$$e^{i\pi(\theta(\hat p+k)-\theta(\hat p))},\eqno(B-2)$$
can be removed from the cocycle operator by a suitable unitary
transformation because the cocycle operator $C_k$ appears always in the
combination $e^{ik\cdot\hat x}C_k$.

The cocycle operator $C_k$ will consist of the zero modes. Since the
vertex operator (3-1) should represent the emission of a state with the
momentum $k^A$, the cocycle operator $C_k$ will not depend on $\hat x^A$
and be represented in terms of $\hat p^A$ as well as $k^A$. We may write
the cocycle operator $C_k$ into the form
$$C_k=e^{i\pi k^AM^{AB}\hat p^B+i\pi F_k(\hat p)},\eqno(B-3)$$
where $M^{AB}$ is defined by eq.(3-9). We require that the zero mode part
of the vertex operator (3-5) satisfies
$$\eqalignno{
V_0(k)V_0(k')&=\varepsilon(k,k')V_0(k+k')&(B-4)\cr
             &=(-1)^{k\cdot\eta k'}V_0(k')V_0(k),&(B-5)\cr}$$
where the phase factor $\varepsilon(k,k')$ is assumed to be c-number. The
above two conditions can be replaced by
$$F_{k+k'}(\hat p)-F_k(\hat p+k')-F_{k'}(\hat p)=\hat p{\rm -independent
\quad mod} \ 2,$$
$$F_k(\hat p+k')+F_{k'}(\hat p)-F_{k'}(\hat p+k)-F_k(\hat p)=0 \quad {\rm
mod} \ 2. \eqno(B-6)$$
It will be convenient to use the following basis of the momentum: Let
$e^A_a$ $(a=1,\cdots,2D)$ be a basis of $\Gamma^{D,D}$, i.e., any momentum
$k^A\in\Gamma^{D,D}$ can be expressed as
$$k^A=\sum_{a=1}^{2D} k^ae^A_a,\quad k^a\in {\bf Z}. \eqno(B-7)$$
In this basis, we assume that $F_k(\hat p)$ can be expanded in powers of
$k^a$ and $\hat p^a$ as follows:
$$F_k(\hat p)=\sum_{n=2}^N\sum_{a_1,\cdots,a_n=1}^{2D}\sum_{j=1}^{n-1}{1
\over j!(n-j)!}\Delta {M_{a_1\cdots a_j}}^{a_{j+1}\cdots a_n}k^{a_1}\cdots
k^{a_j}\hat p^{a_{j+1}}\cdots\hat p^{a_n},$$
$$ \qquad\qquad\qquad\qquad\qquad (k^a,\hat p^a\in {\bf Z}),\eqno(B-8)$$
where $N$ is an arbitrary positive integer and the coefficient $\Delta
{M_{a_1\cdots a_j}}^{a_{j+1}\cdots a_n}$ is totally symmetric with respect
to lower indices or upper indices.

Our aim of this appendix is now to show that by a suitable unitary
transformation $F_k(\hat p)$ can always reduce to
$$F_k(\hat p)=0 \quad {\rm mod} \ 2, \eqno(B-9)$$
up to $\hat p$-independent constant terms. Before we prove eq.(B-9) for
arbitrary $N$, it may be instructive to examine the case of $N=3$, i.e.,
$$F_k(\hat p)=\sum_{a,b}\Delta {M_a}^bk^a\hat p^b+\sum_{a,b,c}{1\over2!}\{
\Delta {M_a}^{bc}k^a \hat p^b\hat p^c +\Delta {M_{ab}}^{c}k^a k^b\hat p^c
\}.\eqno(B-10)$$
Inserting eq.(B-10) into eqs.(B-6) and comparing the third order terms of
both sides of eqs.(B-6) with respect to $k^a,k'^a$ and $\hat p^a$, we find
$$\sum_{a,b,c}\{\Delta {M_a}^{bc}k^ak'^b\hat p^c-\Delta {M_{ab}}^c k^ak'^b
\hat p^c\}=0 \quad {\rm mod} \ 2,$$
$$\sum_{a,b,c}{1\over2!}\{\Delta {M_a}^{bc}k^ak'^bk'^c-\Delta
{M_{ab}}^ck'^ak'^b k^c\}=0 \quad {\rm mod} \ 2,\eqno(B-11)$$
for all $k^a,k'^a,\hat p^a\in{\bf Z}$. From these equations, we can show
the following equality:
$$\eqalignno{
&\sum_{a,b,c}{1\over2!}\{\Delta {M_a}^{bc}k^a\hat p^b\hat p^c+\Delta
{M_{ab}}^{c}k^ak^b\hat p^c \}&\cr
&=\sum_{a=b=c}{1\over3!}\Delta {M_a}^{aa}\{(\hat p^a+k^a)^3-(\hat p^a)^3-
(k^a)^3\}&\cr
&+(\sum_{a=b<c}+\sum_{c<a=b}){1\over2!}\Delta {M_c}^{aa}\{(\hat p^a+k^a)^2
(\hat p^c+k^c)-(\hat p^a)^2\hat p^c-(k^a)^2k^c\}&\cr
&+\sum_{a<b<c}\Delta {M_a}^{bc}\{(\hat p^a+k^a)(\hat p^b+k^b)(\hat p^c
+k^c)-\hat p^a\hat p^b\hat p^c-k^ak^bk^c\} \quad{\rm mod} \ 2.\qquad\qquad
&(B-12)}$$
Then, it follows that the second and the third terms in the right hand
side of eq.(B-10) can be removed by a suitable unitary transformation (up
to $\hat p$-independent terms). Next comparing the second order terms of
both sides of eqs.(B-6) with respect to $k^a,k'^a$ and $p^a$, we find
$$\sum_{a,b}\{\Delta {M_a}^bk^ak'^b-\Delta {M_a}^bk'^ak^b\}=0 \quad {\rm
mod} \ 2. \eqno(B-13)$$
Without loss of generality, we can assume that the matrix $\Delta {M_a}^b$
is antisymmetric because the symmetric part of $\Delta {M_a}^b$ can be
removed by a suitable unitary transformation. Since eq.(B-13) then means
that $\Delta {M_a}^b\in {\bf Z}$, we can introduce a symmetric matrix
$S_{ab}$ through the relation
$$S_{ab}=\Delta {M_a}^b \quad {\rm mod} \ 2.\eqno(B-14)$$
In terms of $S_{ab}$, the first term of eq.(B-10) can be written as
$$\eqalignno{
\sum_{a,b}\Delta {M_a}^bk^a\hat p^b&=\sum_{a,b}S_{ab}k^a\hat p^b \quad {
\rm mod} \ 2&\cr
&=\sum_{a,b}{1\over2!}S_{ab}\{(\hat p^a+k^a)(\hat p^b+k^b)-\hat p^a\hat
p^b-k^ak^b\} \quad {\rm mod} \ 2.\qquad &(B-15)\cr}$$
Therefore, we have proved that by suitable unitary transformations $F_k(
\hat p)$ given in eq.(B-10) can reduce to eq.(B-9) up to $\hat
p$-independent terms.

Let us prove eq.(B-9) for arbitrary $N$. By inserting eq.(B-8) into eqs.
(B-6) and by comparing the $N$th order terms of both sides of eqs.(B-6)
with respect to $k^a,k'^a$ and $\hat p^a$, it is not difficult to show the
following equality:
$$\eqalignno{
&\sum_{a_1,\cdots,a_N}\sum_{j=1}^{N-1}{1\over j!(N-j)!}\Delta {M_{a_1
\cdots a_j}}^{a_{j+1}\cdots a_N}k^{a_1}\cdots k^{a_j}\hat p^{a_{j+1}}
\cdots\hat p^{a_N} & \cr
&=\sum_a\{{1\over 2!(N-2)!}(\Delta {M_{aa}}^{a\cdots a}-\Delta {M_a}^{a
\cdots a})(k^a)^2(\hat p^a)^{N-2}&\cr
&+{1\over 3!(N-3)!}(\Delta {M_{aaa}}^{a\cdots a}-\Delta {M_a}^{a\cdots a})
(k^a)^3(\hat p^a)^{N-3}&\cr
&+\cdots+{1\over (N-2)!2!}(\Delta {M_{a\cdots a}}^{aa}-\Delta {M_a}^{a
\cdots a})(k^a)^{N-2}(\hat p^a)^2\} \quad {\rm mod} \ 2,&(B-16)\cr}$$
with
$$\Delta {M_{\underbrace{\scriptstyle{a\cdots a}}_{j}}}^{\overbrace{
\scriptstyle{a\cdots a}}^{N-j}}-\Delta {M_a}^{\overbrace{\scriptstyle{a
\cdots a}}^{N-1}}=0 \quad {\rm mod} \ 2(j-1)!(N-j)!,\eqno(B-17)$$
where in the right hand side of eq.(B-16) we have omitted $\hat
p$-independent terms as well as the terms which can be removed by unitary
transformations. It follows from the theorem (A-1) that all the terms in
the right hand side of eq.(B-16) can reduce to lower order terms with
respect to $k^a$ and $\hat p^a$ and hence they can be absorbed into lower
order terms of $F_k(\hat p)$ by suitably redefining the coefficients of
the lower order terms of $F_k(\hat p)$. Therefore, we conclude that the
$N$th order terms of $F_k(\hat p)$ can be put to be equal to zero.

Repeating the above arguments order by order, we finally come to the
conclusion (B-9).

\endpage

\references

\item{[1]} L. Dixon, J.A. Harvey, C. Vafa and E. Witten, Nucl. Phys. {\bf
B261} (1985) 678; {\bf B274} (1986) 285.
\item{[2]} H. Kawai, D. Lewellyn and A.H. Tye, Phys. Rev. Lett. {\bf 57}
(1986) 1832; Nucl. Phys. {\bf B288} (1987) 1;
\item{   } I. Antoniadis, C. Bachas and C. Kounnas, Nucl. Phys. {\bf B289}
(1987) 87.
\item{[3]} W. Lerche, A.N. Schellenkens and N.P. Warner, Phys. Rep. {\bf
177} (1989) 1.
\item{[4]} D. Gepner, Phys. Lett. {\bf B199} (1987) 380; Nucl. Phys. {\bf
B296} (1987) 757.
\item{[5]} Y. Kazama and H. Suzuki, Nucl. Phys. {\bf B321} (1989) 232.
\item{[6]} C. Vafa and N.P. Warner, Phys. Lett. {\bf B218} (1989) 51;
\item{    } W. Lerche, C. Vafa and N.P. Warner, Nucl. Phys. {\bf B324}
(1989) 427;
\item{    } P.S. Howe and P.C. West, Phys. Lett. {\bf B223} (1989) 377; {
\bf B244} (1989) 270.
\item{[7]} E.S. Fradkin and A.A. Tseytlin, Phys. Lett. {\bf B158} (1985)
316; Nucl. Phys. {\bf B261} (1985) 1;
\item{    } C.G. Callan, D. Friedan, E.J. Martinec and M.J. Perry, Nucl.
Phys. {\bf B262} (1985) 593;
\item{    } C.G. Callan, I.R. Klebanov and M.J. Perry, Nucl. Phys. {\bf
B278} (1986) 78;
\item{    } T. Banks, D. Nemeschansky and A. Sen, Nucl. Phys. {\bf B277}
(1986) 67.
\item{[8]} P. Candelas, G. Horowitz, A. Strominger and E. Witten, Nucl.
Phys. {\bf B258} (1985) 46.
\item{[9]} A. Font, L.E. Ib\'a\~ nez, F. Quevedo and A. Sierra, Nucl.
Phys. {\bf B331} (1990) 421.
\item{[10]} J.A. Casas and C. Mu\~ noz, Nucl. Phys. {\bf B332} (1990) 189.
\item{[11]} Y. Katsuki, Y. Kawamura, T. Kobayashi, N. Ohtsubo, Y. Ono and
K. Tanioka, Nucl. Phys. {\bf B341} (1990) 611.
\item{[12]} A. Fujitsu, T. Kitazoe, M. Tabuse and H. Nishimura, Intern. J.
Mod. Phys. {\bf A5} (1990) 1529.
\item{[13]} K.S. Narain, M.H. Sarmadi and C. Vafa, Nucl. Phys. {\bf B288}
(1987) 551; {\bf B356} (1991) 163.
\item{[14]} K.S. Narain, Phys. Lett. {\bf B169} (1986) 41;
\item{    } K.S. Narain, M.H. Sarmadi and E. Witten, Nucl. Phys. {\bf
B279} (1987) 369.
\item{[15]} J.H. Schwarz, Phys. Rep. {\bf 8C} (1973) 269; {\bf 89} (1982)
223;
\item{    } J. Scherk, Rev. Mod. Phys. {\bf 47} (1975) 123.
\item{[16]} D.J. Gross, J.A. Harvey, E. Martinec and R. Rohm, Nucl. Phys.
{\bf B256} (1985) 253; {\bf B267} (1986) 75.
\item{[17]} J.L Cardy, Nucl. Phys. {\bf B270} (1986) 186.
\item{[18]} H. Suzuki and A. Sugamoto, Phys. Rev. Lett. {\bf 57} (1986)
1665.
\item{[19]} N. Sakai and Y. Tanii, Nucl. Phys. {\bf B287} (1987) 457.
\item{[20]} K. Inoue, S. Nima and H. Takano, Prog. Theor. Phys. {\bf 80}
(1988) 881.
\item{[21]} A.M. Polyakov, Phys. Lett. {\bf B103} (1981) 207; {\bf B103}
(1981) 211.
\item{[22]} I. Frenkel and V. Ka\v c, Invent. Math. {\bf 62} (1980) 23;
\item{    } G. Segal, Commun. Math. Phys. {\bf 80} (1981) 301;
\item{    } P. Goddard and D. Olive, Intern. J. Mod. Phys. {\bf A1} (1986)
303.
\item{[23]} V.A. Kosteleck\'y, O. Lechtenfeld, W. Lerche, S. Samuel and S.
Watamura, Nucl. Phys. {\bf B288} (1987) 173.
\item{[24]} M. Sakamoto, Phys. Lett. {\bf B231} (1989) 258.
\item{[25]} C. Vafa, Nucl. Phys. {\bf B273} (1986) 592.
\item{[26]} T. Horiguchi, M. Sakamoto and M. Tabuse, Kobe preprint
KOBE-92-03 (1992).
\item{[27]} M. Sakamoto and M. Tabuse, Phys. Lett. {\bf B260} (1991) 70.
\item{[28]} M. Sakamoto, Prog. Theor. Phys. {\bf 84} (1990) 351.
\item{[29]} K. Itoh, M. Kato, H. Kunitomo and M. Sakamoto, Nucl. Phys. {
\bf B306} (1988) 362.
\item{[30]} J. Erler, D. Jungnickel, J. Lauer and J. Mas, preprint
SLAC-PUB-5602 (1991).

\end